\documentclass[review]{elsarticle}
\usepackage{amsmath,amssymb}
\usepackage{bbm}
\usepackage{graphicx,graphics}
\usepackage{lineno,hyperref}
\usepackage{soul}
\usepackage[makeroom]{cancel}
\usepackage{comment}
\usepackage[normalem]{ulem} 
\usepackage{xcolor} 

\modulolinenumbers[1]

\journal{Elsevier}

\begin{document}

\begin{frontmatter}

\title{Assessing the value of energy storage systems for distribution grid applications}

\author[mymainaddress1]{Sahar Moghimian Hoosh}
\cortext[mycorrespondingauthor]{Corresponding author}
\ead{sahar.moghimian@skoltech.ru}

\author[mymainaddress1]{H. Ouerdane}

\author[mymainaddress2]{Vladimir Terzija}

\author{David Pozo}

\address[mymainaddress1]{Center for Digital Engineering, Skolkovo Institute of Science and Technology, Moscow 121205, Russia}
\address[mymainaddress2]{School of Engineering, Merz Court E4.41, Newcastle University, Newcastle upon Tyne NE1 7RU, United Kingdom}

\begin{abstract}

We analyze the potential benefits that energy storage systems (ESS) can bring to distribution networks in terms of cost, stability and flexibility. We propose an optimization model for the optimal sizing, siting, and operation of storage systems in distribution grids. A DistFlow formulation is used for modeling the AC power flow. The ESS model is based on a generic formulation that captures the charging and discharging modes' complementarity. The resulting optimization model is stated as a mixed-integer quadratically constrained program (MIQCP) problem. The optimization model is assessed on the modified 33-bus IEEE network, which includes renewable energy resources and ESS. The obtained results show that ESS can offer various important benefits such as overall cost reduction, energy arbitrage, voltage regulation, and congestion management in distribution grids. These findings highlight the significance of utilizing ESS technologies to provide aggregated value through various grid services, extending beyond energy arbitrage alone.
\end{abstract}


\end{frontmatter}


\section{Introduction}
High integration of renewable energy sources in power systems poses challenges to their normal operation \cite{EVANS20124141}. Power quality issues, voltage fluctuation, and system stability are some of the problems caused by the intermittent nature of variable sources. In such cases, traditional approaches involve the replacement or upgrading of transformers, conductors, and overhead lines with new equipment. However, more flexible solutions can expedite the process of grid modernization \cite{8403356}. Energy storage systems (ESS) have shown potential as multifaceted technologies for enhancing grid performance.
Optimal utilization of ESS in power grids can yield economic profits and support various power system applications, including power quality improvement, peak shaving, demand response, voltage and frequency regulation, and infrastructure upgrade deferral \cite{YEKINISUBERU2014499}. Despite extensive studies on the significant role of storage systems in power grids \cite{8000386,7968249, Alhamalii, TAN2021102591}, questions remain regarding the techno-economic value of ESS. The technological value of energy storage devices is primarily assessed based on the services they provide to different sectors of energy systems. Multipurpose ESS models, capable of simultaneously providing several grid services, have the potential to significantly increase economic benefits for investors and owners. To develop such models, it is necessary to determine the optimal size, location, and operation of storage devices \cite{xiao2016determination}. Various optimization approaches have been developed for this purpose, including analytical approaches, mathematical programming (MP), heuristic techniques, meta-heuristics, and hybrid methods \cite{PRAKASH2016111}.

In this research, we focus on MP as the primary concern is guaranteeing the optimal solution rather than computation time. ESS optimization problems in MP are typically formulated as linear, nonlinear, or mixed-integer programming models, and they can be solved using different techniques such as second-order cone programming \cite{GROVERSILVA2018385,qin2016optimal,nick2014optimal}, dynamic programming (DP) \cite{en12061098,en9060462, 7079529,sui2014energy,rurgladdapan2012li}, and stochastic programming (SP) \cite{7592939}.                            
In \cite{7540850}, a linearized multi-period optimal power flow method is applied for ESS allocation and solved using the forward-backward sweep power flow with power loss reduction application. A convex formulation of the AC optimal power flow (AC-OPF) is used in \cite{nick2014optimal} to determine the optimal location and capacity of the ESS in the network for voltage support, load curtailment, and congestion management. Additionally, a stochastic mixed-integer linear programming method proposed in \cite{article6} aims to minimize the operating and investment costs of the storage system in transmission networks while providing an energy arbitrage service. However, most of these studies simplify the mathematical setting of the problem to enhance tractability or reduce computational cost at the expense of the accuracy. 
\\
We thus aim to propose an ESS planning framework that (i) guarantees the profitable operation of ESS units and (ii) provides a comprehensive mathematical suite for the optimal allocation of the ESS without oversimplifying the nonlinear and discrete nature of the problem. The main contributions of our work are ESS modeling, and assessment of ESS value for the provision of several grid services in distribution networks.
\\
The article is organized as follows: section \ref{1.Math} develops the mathematical model of the problem. Section \ref{3.TestResults} presents our case study and optimization results. We conclude the article with final remarks and an outline of future research directions in section \ref{4.conclude}.
\\
\section{Mathematical Formulation}
 \label{1.Math}
The proposed mathematical model is DistFlow \cite{6760226} to which we add the ESS constraints. The DistFlow model is shown to be equivalent to the bus injection AC power flow model, providing confidence in its accuracy \cite{6897933}.
For larger networks, the computational performance of the DistFlow model is more efficient since the non-convex nature of the AC-OPF makes it challenging to achieve the globally optimal solution \cite{25627, GRANGEREAU2022107774}.

The investment problem is formulated based on a specified target year. To capture the dynamic nature of supply and demand fluctuations throughout the year, we have selected twelve representative days (indexed by $\omega$). The annual investment costs are then updated on daily values. Additionally, the weights assigned to each representative day are normalized to reflect the probabilities of their occurrence, denoted as $\rho_{\omega}$. This approach allows us to consider the different scenarios of the photovoltaic (PV) system output and its occurrence probabilities within the time period interval of one hour. Sets in this formulation are described as follows:
\begin{eqnarray*}
N &:=& \{1,…,m_{n}\} ~ \mbox{set of indices of buses} \\
L &:=& \{1,…,l_{n}\}~ \mbox{set of indices of branches}\\
G &:=& \{1,…,g_{n}\}~ \mbox{set of indices of generators}\\
S &:=& \{1,…,\omega_{n}\}~ \mbox{set of indices of scenarios}\\
T &:=& \{1,…,t_{n}\}~ \mbox{set of indices of time intervals}
\end{eqnarray*}

\subsection{Objective Function}
We introduce the objective function:

\begin{equation}
\min{C} = \min{ \big( C_{i,t,\omega}^{Gen} + C_{inv}^{ESS} + C_{opt,t,\omega}^{ESS} \big)}  \label{3.1}
\end{equation}
\noindent where $C_{i,t,\omega}^{Gen}$ represents the quadratic cost of small fuel-based generators \cite{articlequad}, $C_{inv}^{ESS}$ denotes the initial investment cost of the ESS, and $C_{opt,t,\omega}^{ESS}$ corresponds to the operational cost of the ESS for a given time period $t$ and scenario $\omega$. The expressions for $C_{i,t,\omega}^{Gen}$, $C_{inv}^{ESS}$, and $C_{opt,t,\omega}^{ESS}$ are given by: 

\begin{equation}
\begin{split}
   C_{i,t,\omega}^{Gen} = \sum_{\omega\in S} \rho_{\omega} \:\sum_{t\in T} &\sum_{i\in G}  a_{i} +  b_{i} p_{i}^{G}(t,\omega) +  c_{i} p_{i}^{G}(t,\omega)^2 \label{eq_Pm}
\\
\end{split}
\end{equation}
\begin{equation}
   \label{3.3}
    C_{inv}^{ESS} = f E_{i}^{ESS}\\   
\end{equation}
\begin{equation}
\label{eq4}
C_{opt,t,\omega}^{ESS} = h(\sum_{\omega\in S}\sum_{t\in T} p_{i}^{ch}(t,\omega) + p_{i}^{dis}(t,\omega))
\end{equation}

\noindent where $a_{i}$, $b_{i}$, and $c_{i}$ are the cost coefficients of power generator unit $i$, which we derived following the approach used in \cite{article9}. The quantities $p_{i}^{G}(t,\omega)$ and $q_{i}^{G}(t,\omega)$ denote the active and reactive power of generator $i$ at time $t$ within scenario $\omega$. $E_{i}^{ESS}$ represents the energy capacity of the storage device, and $f$ represents the ESS investment cost per unit capacity, which in this work is set to 250 ($\$$/kWh) \cite{8602146}. The variables $p_i^{ch}(t,\omega)$ and $p_i^{dis}(t,\omega)$ refer to the ESS charging and discharging power over scenario $\omega$ and time period $t$, respectively. The parameter $h$ represents the ESS operational cost, which is equal to 0.005 ($\$$/kWh) \cite{8602146}.

\subsection{Network Constraints (DistFlow Model)}

\paragraph{Branch Power Flow:} To linearize the branch flow equations, the voltage of bus $i$, $V_{i}(t,\omega)$, is replaced by $w_{i}(t,\omega)$, and $I_{ij}(t,\omega)$, the current of the branch $(i,j)$ is replaced by $l_{ij}(t,\omega)$ as the auxiliary variables: 
\begin{equation}
     w_{i}(t,\omega) = V_{i}(t,\omega)^{2}
    \label{3.6}   
\end{equation} 
\begin{equation}
    \label{eq6.} 
    l_{ij}(t,\omega) = I_{ij}(t,\omega)^{2}  
\end{equation}

\noindent so that the Distflow equations are given by: 

\begin{equation}
    \label{3.7}
    p_{ij}(t,\omega) = p_{j}(t,\omega) + r_{ij}l_{ij}(t,\omega) + \sum_{k:(j,k)\in L}p_{jk}(t,\omega) \\
\end{equation}
\begin{equation}
    \label{3.8}
    q_{ij}(t,\omega) = q_{j}(t,\omega) + x_{ij}l_{ij}(t,\omega) + \sum_{k:(j,k)\in L}q_{jk}(t,\omega) \\ 
\end{equation}
\begin{equation}
\begin{split}
 w_{j}(t,\omega) =& w_{i}(t,\omega) + (r_{ij}^2+x_{ij}^2)\ l_{ij}(t,\omega)\\&-2(r_{ij}p_{ij}(t,\omega)+x_{ij}q_{ij}(t,\omega)) \label{3.9}
\end{split}
\end{equation} 
\begin{equation}
p_{ij}(t,\omega)^2+q_{ij}(t,\omega)^2 = l_{ij}(t,\omega)w_{i}(t, \omega)\label{3.10}
\end{equation}

\noindent where $p_{j}(t,\omega)$ and $q_{j}(t,\omega)$ represent the net active and reactive power withdrawn at bus $j$. The line impedance is modeled by $r_{ij}+ jx_{ij}$, where $r_{ij}$ denotes the resistance and $x_{ij}$ represents the reactance. Here, $j$ denotes the imaginary unit with the property $j^2 = -1$. The quantities $p_{ij}(t,\omega)$ and $q_{ij}(t,\omega)$ correspond to the active and reactive branch power flow from node $i$ to $j$, while $p_{jk}(t,\omega)$ and $q_{jk}(t,\omega)$ refer to the active and reactive branch power flow from node $j$ to $k$, respectively.
\paragraph{Nodal Balance:}~

\begin{equation}
 \begin{split}
 p_{i}(t,\omega) = p_{i}^D(t,\omega) -p_{i}^G(t,\omega) - p^{PV}(t,\omega) \\  - p_{i}^{dis}(t,\omega) +p_{i}^{ch}(t,\omega)  \label{3.11}\\
 \end{split}   
\end{equation}
\begin{equation}
 \begin{split}
   q_{i}(t,\omega) = q_{i}^D(t,\omega)-q_{i}^G(t,\omega)+ q_{i}^{inv}(t,\omega)\label{3.12}  
 \end{split}
\end{equation}

\noindent where $p_{i}^D(t,\omega)$ and $q_{i}^D(t,\omega)$ are the active and reactive power consumed by the load located at bus $i$, and $p^{PV}(t,\omega)$ is the generated active power by the PV system. Here, it must be noted that we assumed the centralized storage device has a built-in smart inverter which can absorb/deliver reactive power from/to the main grid. So, the reactive power of the built-in inverter, $q_{i}^{inv}(t,\omega)$, can take both positive and negative values. The transferred reactive power is used for voltage regulation purposes.

\paragraph*{ Technical Limits:}
The operational limits for the specific generator at bus $i$ are given by:

\begin{eqnarray}
    p_{i}^{G,min}\leq &p_{i}^G(t,\omega)& \leq p_{i}^{G,max}
    \label{3.13}\\
    q_{i}^{G,min} \leq &q_{i}^G(t,\omega)& \leq q_{i}^{G,max} \label{3.14}
\end{eqnarray}    

\noindent where $p_{i}^{G,min}$ and $p_{i}^{G,max}$ are the minimum and maximum active power output limits, and $q_{i}^{G,min}$ and $q_{i}^{G,max}$ are the minimum and maximum reactive power output limits, respectively.

\paragraph*{Thermal Limit of Lines:}
\begin{equation}\label{3.16}
p_{ij}^2(t,\omega) + q_{ij}^2(t,\omega) \leq (S_{ij}^{max})^2
\end{equation}

\paragraph*{Bus Voltage Limits:}

\begin{equation}\label{3.17}
w_{i}^{min} \leq w_{i}(t,\omega) \leq w_{i}^{max}
\end{equation}

\noindent where $w_{i}^{min}$ and $w_{i}^{max}$ are lower and upper limits of the bus voltage, and $ S_{ij}^{max}$ is the maximum apparent power of the line connecting the bus $i$ to $j$.

\paragraph*{Substation Boundary Conditions:}
\begin{eqnarray}
    \label{3.19}
    w_{1} &=& 1.0 \quad 
\end{eqnarray}

\noindent where $w_1$ is the voltage of the slack bus.

\subsection{ESS Constraints} 
The ESS is modeled as an ideal and generic storage device \cite{6783800} with the inclusion of binary variables that prevent simultaneous charging and discharging operations. Previous research highlighted in \cite{Pozo_PSCC_2022} has demonstrated that the complementarity of charging and discharging is often violated in power grids operating at their limits. To address this issue, we employ two binary status variables, namely $x_{i}^{ch}(t,\omega)$ and $x_{i}^{dis}(t,\omega)$, to capture the charging and discharging of the ESS located at bus $i$. The following constraints ensure that simultaneous charging and discharging are not allowed for each time period $t$, and scenario $\omega$. 
{\small\begin{equation}
x_{i}^{ch}(t,\omega) + x_{i}^{dis} (t,\omega) \leq 1,  \quad\quad x_{i}^{ch}(t,\omega),x_{i}^{dis}(t,\omega) \in \{0,1\}
\label{3.28}
\end{equation}}

The operation of the ESS is limited to the rated active and reactive power and the state of energy (SoE). The active and reactive power limitations for ESS and its power converter are expressed as:  

\begin{equation}
0 \leq p_{i}^{ch}(t,\omega) \leq p_i^{ch,max} x_{i}^{ch}(t,\omega) 
\label{3.22}
\end{equation}
\begin{equation}
0 \leq p_{i}^{dis}(t,\omega) \leq p_i^{dis,max}  x_{i}^{dis}(t,\omega)
\label{3.23}
\end{equation}
\begin{equation}
- q_i^{inv,min} \leq q_{i}^{inv}(t,\omega) \leq q_i^{inv,max} 
\label{3.24}
\end{equation}

\noindent where $p_i^{ch,max}$ and $p_i^{dis,max}$ represent the maximum permissible charge and discharge power of ESS $i$, while $q_i^{inv,min}$ and $q_i^{inv,max}$ denote the absolute lower and upper limits of the inverter's reactive power, respectively. The state of energy (SoE) is represented by $e_{i}(t,\omega)$.  It is commonly expressed as a percentage, where 0\% represents no energy stored or an empty state, and 100\% represents the maximum energy capacity or a fully charged state. The SoE inventory at time $t+1$ is determined by: 

{\small\begin{equation}
e_{i}({t+1},\omega) = e_{i}(t,\omega) + \eta_i^{ch} p_{i}^{ch}(t,\omega) - \frac{1}{\eta_i^{dis}}  p_{i}^{dis}(t,\omega)
\label{3.29}
\end{equation}}

\noindent where $\eta_i^{dis}$ is ESS's discharging efficiency, and $\eta_i^{ch}$  is the charging efficiency.

To avoid 'border effects', we assume that energy storage at the end of the period is equal to the initial energy storage: $e_{i}(1) = e_{i}(25)$. This is a common approach followed in the literature to avoid the potential where the ESS may end up being empty in the last period of the optimization horizon \cite{8927263}.

Finally, the limit on the ESS energy capacity investment is given by:

{\small\begin{equation}\
E_i^{ESS} \leq E_i^{max} y_{i}
\label{3.27}
\end{equation}}

\noindent where $ E_i^{max} $ is the maximum energy capacity of the storage device that is available for investment, and $y_{i}$ is the binary variable that represents the location of the centralized ESS.

The resulting optimization problem \eqref{3.1} -- \eqref{3.27} corresponds to a nonlinear, and non-convex model with respect to the discrete and continuous ESS variables and bilinear products in \eqref{3.10}. However, by relaxing \eqref{3.10} with a cone constraint \cite{6760226}, the problem can be formulated as a mixed integer quadratically constrained program (MIQCP), and easily solved using commercial solvers like Gurobi.

\section{Numerical Analysis} \label{3.TestResults}

\subsection{Test System}
The proposed model is validated using a modified version of the IEEE 33 bus system, which has been previously described in \cite{9258930}. The modified system includes a single feeder substation, four distributed generation (DG) units, and two reactive power compensators (RPCs) located at bus 18 and bus 33 to provide voltage support. The specific locations of the RPCs are obtained from the data input provided in \cite{9258930}. Additionally, we introduced a PV generator located at bus 4. Historical data are used for variables, including PV generation output, and load profile \cite{8602146}. The PV generation is represented by twelve generation scenarios, with four scenarios assigned to each season, and three days per season to capture the variability of solar irradiance and weather conditions. All twelve scenarios have the same probability. The model is constructed using the JuMP package for Julia language and solved with Gurobi by a laptop computer with an Intel(R) Core(TM) i7-11370H @3.30GHz and 16GB of RAM. 

\subsection{Integration of ESS in the Modified IEEE 33-bus System}

\begin{figure}
    \centering
    \includegraphics[scale=0.14]{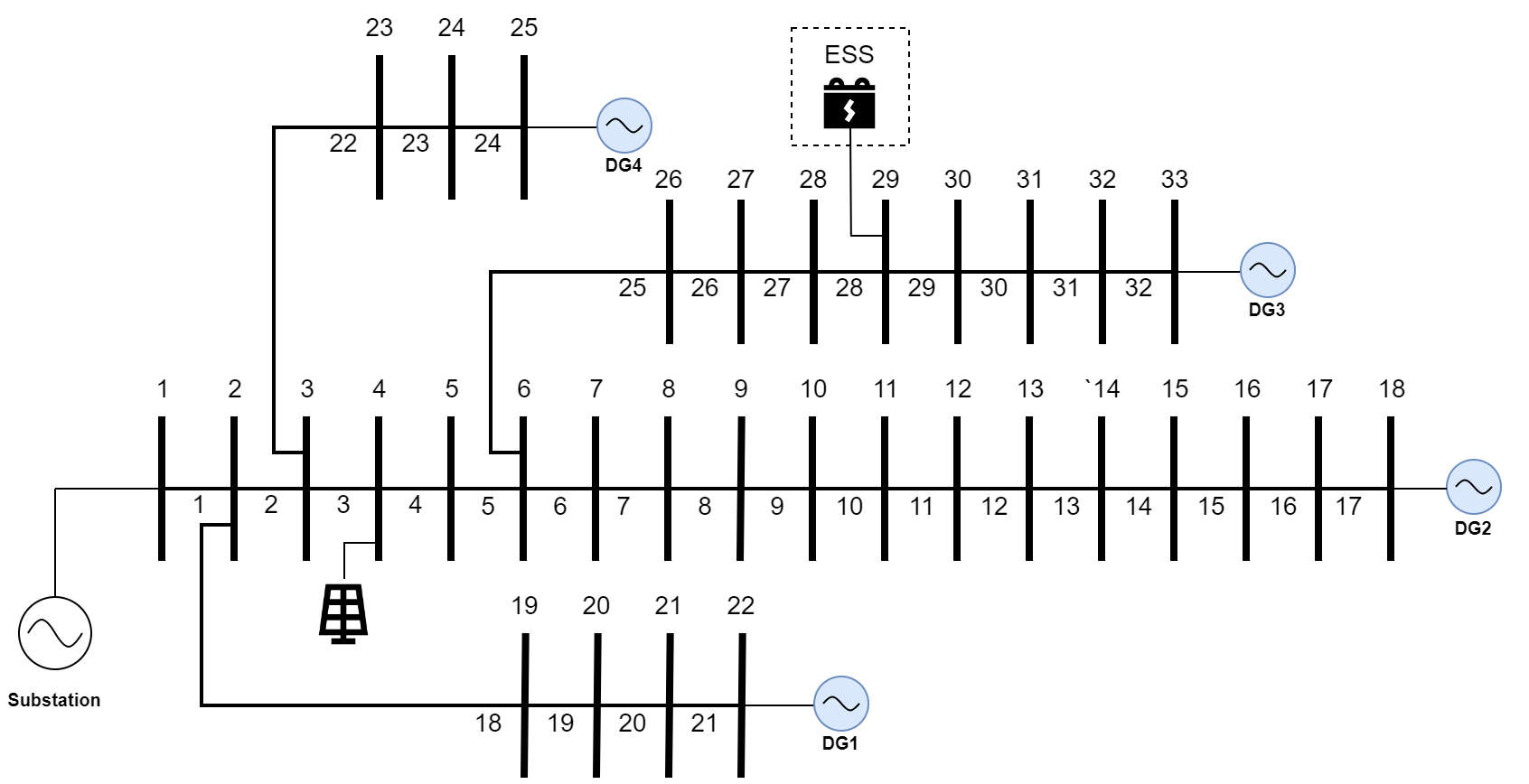}
     \caption{Modified IEEE 33-bus system test case with allocated ESS.}
    \label{fig:4.9.2}
\end{figure}

The original network configuration obtained from \cite{9258930} revealed indications of an ill-conditioned network, particularly in terms of voltage regulation, resulting in an infeasible scenario for the optimal power flow (OPF) problem. To mitigate this challenge, RPCs were utilized for nodal voltage regulation. However, in this study, our focus shifts towards exploring an alternative solution. Specifically, we aim to demonstrate the effectiveness of an ESS in replacing RPCs, offering potential cost savings, as well as providing grid-related benefits. The application of our optimization model to determine the optimal size and location of the integrated ESS shows that a centralized ESS with an energy capacity of 710 (kWh) should be allocated to bus 29 as depicted in Fig.\ref{fig:4.9.2}. The technical parameters of the ESS can be found in Table \protect\ref{Table4.6}, with all values expressed in per unit (pu).

\begin{table}
\centering
\caption{ESS parameters.}\label{Table4.6}
\begin{tabular}{ccccc}
\hline
\hline
\textbf{$E^{max}$}[pu] & \textbf{$p^{ch,max}$}[pu] & \textbf{$p^{dis,max}$}[pu] &  \textbf{$\eta^{ch}$} & \textbf{$\eta^{dis}$} \\
1.2 & 2 & 2 & 85\% & 90\% \\
\hline
\hline
\end{tabular}
\end{table}

Assessing the total cost of the system under three conditions: 
\begin{enumerate}[i)]
\item The absence of voltage regulators;
\item The use of RPCs;
\item The integration of a centralized ESS at bus 29;
\end{enumerate}

the results of the OPF analysis presented in Table \ref{table:4.9} demonstrate a noticeable cost reduction in the system with ESS. This highlights the potential of the ESS as a viable alternative to traditional voltage regulators, resulting in improved performance and cost savings for the distribution network. In contrast to RPCs, which have limited grid functionality, an ESS can efficiently provide multiple grid services at the same time. Though it takes more time to
solve the problem with ESS, it is still acceptable for planning
models in power systems.

\begin{table}
\centering
\caption{OPF results.}\label{table:4.9}
  \begin{tabular}{cccccc}
    \hline
		\hline
    Case  & Count & Location & Total Cost [$\$$/hr] & Gap [\%] & Run-time [s]\\
    \hline
    None & -- & -- & Infeasible & Infeasible & Infeasible \\ 
    RPC & 2 & 18, 33 & 1382.19 & 0.0913 & 29.38 \\ 
    ESS & 1 & 29 & 424.72 & 0.0371 & 2126.69 \\  
    \hline
		\hline
  \end{tabular}
\end{table}

\subsection{Results on ESS for Grid Service Provision}
In this section, we present the results of our analysis on the use of ESS for providing services to the grid. We specifically examine how ESS performs in energy arbitrage, voltage regulation, and congestion management.

\subsubsection{Energy Arbitrage}
\begin{figure}[b]
    \centering
    \includegraphics[scale=1]{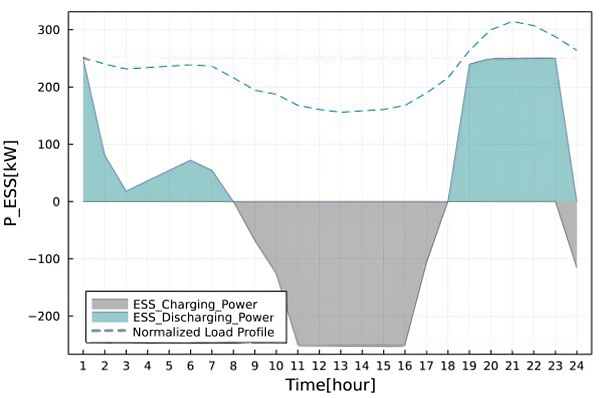}

     \caption{ESS charging and discharging pattern for energy arbitrage purpose.}
    \label{fig:4.9}
\end{figure}

Figure \ref{fig:4.9} shows the optimal profile of the ESS charging and discharging over the course of a single day. It demonstrates the capability of ESS to store energy during low-cost periods when demand is low and discharge it at higher prices during periods of high demand. This mechanism enables the ESS to maximize economic benefits through the exploitation of price differentials.

\subsubsection{Voltage Regulation}
\begin{figure}
    \centering
    \includegraphics[scale=0.6]{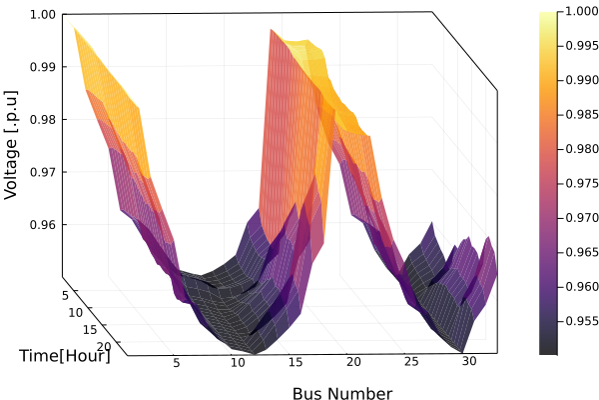}
    \caption[Voltage profile of the modified 33-bus network with RPCs.]{Voltage profile of the modified 33-bus network with RPCs.}
    \label{fig:4.10}
\end{figure}

\begin{figure}
    \centering  \includegraphics[scale=0.6]{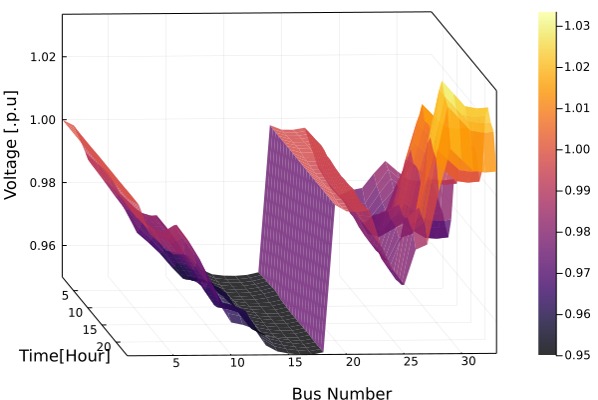}
    \caption{Voltage profile of the modified 33-bus network with an ESS.}
    \label{fig:4.11}
\end{figure}

Here we assess the impact of ESS on enhancing the voltage profile within the system. Figure \ref{fig:4.10} illustrates the voltage profile of the system with RPCs. While the voltage levels of all buses fall within the acceptable limits ($V_{min} = 0.95\ pu, V_{max}= 1.05\ pu$), nodes 12 and 29 exhibit voltages that are in close proximity to the minimum voltage threshold.
Consequently, based on the observations from the figure, we anticipate that the model will identify either node 12 or node 29 as the optimal candidate for ESS installation. Based on the optimization results, our hypothesis is confirmed, as the model designates bus 29 as the most suitable location for implementing an ESS.
Additionally, Fig. \ref{fig:4.11} illustrates the system's voltage profile after optimally replacing RPCs with the storage device. The installation of an ESS at bus 29 effectively eliminates a voltage valley and regulates the voltage profile. This placement significantly reduces the risk of under voltage across neighboring buses. A comparison between the two results demonstrates that an ESS enables more flexible and versatile bi-directional exchange of reactive power.

\subsubsection{Congestion Management}
To analyze the impact of ESS on congestion management, we simulated a stress state in the distribution grid. For this purpose, we increased the load level by 10\% for each hour between 17:00 and 24:00 to identify the critical threshold at which line congestion occurs. Table \ref{table: 4.9} presents the results of OPF analysis for a system with RPCs under different load conditions. It indicates that with an 11\% increase in load level during peak hours, the system becomes infeasible due to line congestion. The congested lines are listed in the table. These findings demonstrate the impact of load increases on the system's feasibility and highlights the presence of congestion issues at higher load levels.

\begin{table}
\centering
\caption{OPF results for the normal and stressed conditions of the network with RPCs.}\label{table: 4.9}
\begin{tabular}{cccccc}
\hline
\hline
\textbf{System with RPCs} & \textbf{\begin{tabular}[c]{@{}c@{}}Normal \\ load level\end{tabular}} & \textbf{\begin{tabular}[c]{@{}c@{}}Load \\ increased \\ by 10\%\end{tabular}} & \textbf{\begin{tabular}[c]{@{}c@{}}Load\\ increased \\ by 11\%\end{tabular}} \\
\hline
\begin{tabular}[c]{@{}c@{}}Total cost\\ $[\$$/hr]\end{tabular} & 1387.9 & 1457.2 & \begin{tabular}[c]{@{}c@{}}Infeasible\end{tabular} \\\\
Congested lines & \multicolumn{3}{c}{1, 2, 3, 4, 5, 7, 22, 25, 26, 27, 28, 29 } \\
\hline
\hline
\end{tabular}
\end{table}

In the next step of the experiment, the test conditions were replicated for the system equipped with an ESS. Remarkably, the optimization problem became infeasible when the load level was increased by 16\% (as indicated in Table \ref{table: 4.10}).

\begin{table}
\centering
\caption{OPF results for the stressed network with an ESS.}\label{table: 4.10}
\begin{tabular}{cccccc}
\hline
\hline
\textbf{System with ESS} & \textbf{\begin{tabular}[c]{@{}c@{}}Normal \\ load level\end{tabular}} & \textbf{\begin{tabular}[c]{@{}c@{}}Load \\ increased \\ by 15\%\end{tabular}} & \textbf{\begin{tabular}[c]{@{}c@{}}Load\\ increased \\ by 16\%\end{tabular}} \\
\hline
\begin{tabular}[c]{@{}c@{}}Total cost\\ $[\$$/hr]\end{tabular} & 424.7 & 608.3 & \begin{tabular}[c]{@{}c@{}}Infeasible\end{tabular} \\\\
Congested lines & \multicolumn{3}{c}{1, 2, 22, 25} \\
\hline
\hline
\end{tabular}
\end{table}

Comparing the data from these tables, it is evident that in the system without storage, a significant number of lines experience congestion during periods of high load. This congestion hampers the flexibility of the network, making it unable to accommodate the increased demand. Consequently, grid enhancements or upgrades are required to address this issue. In contrast, the integration of an ESS proves to be beneficial. It effectively increases the thermal line capacity by at least 5\%, offering a viable solution for utilities to defer overhead line upgrades.

\begin{figure}
    \centering   
    \includegraphics[scale=0.4]{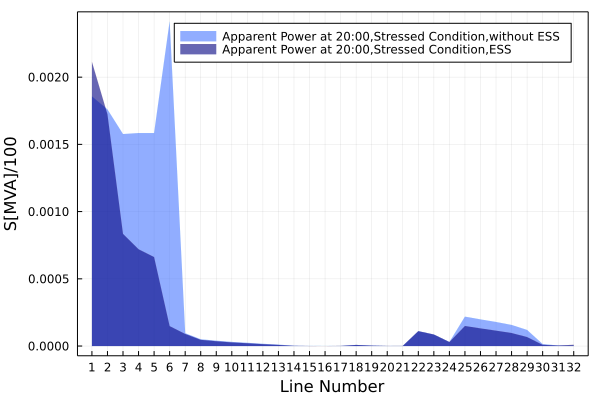}
    \caption{Comparison of apparent power distribution in the network with and without an ESS, under stressed conditions}.
    \label{fig:4.14}
\end{figure}
\begin{figure}[t]
    \centering
    \includegraphics[scale=0.4]{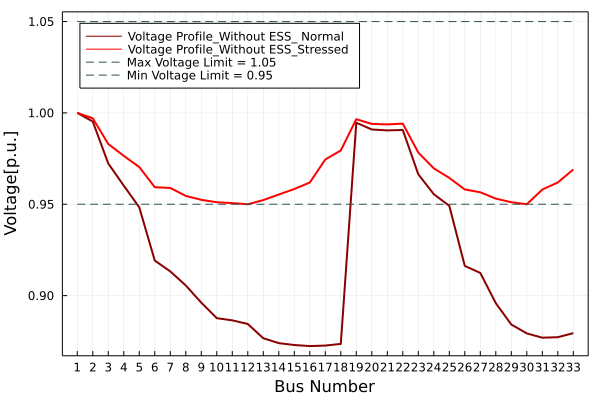}
    \caption {Voltage profile without an ESS: A comparison between congested and normal operating conditions}.
    \label{fig:13}
\end{figure}
\begin{figure}[hbt!]
    \centering
    \includegraphics[scale=0.4]{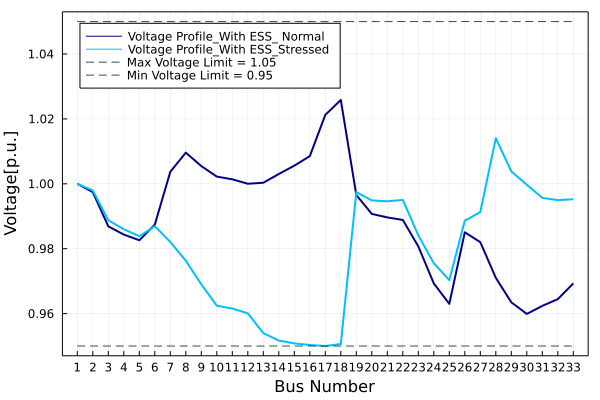}
    \caption{Voltage profile with an ESS: A comparison between congested and normal operating conditions}.
    \label{fig:14}
\end{figure}

Moreover, the distribution of apparent power across different lines shown in Fig.\ref{fig:4.14} highlights any instance of high power flow or potential overloading. This allows to compare the system with and without an ESS under stressed network conditions. The analysis focuses on a specific peak hour at 20:00. In the system without an ESS, there is a noticeable increase in apparent power, particularly on line 7, indicating a potential congestion issue. On the other hand, the integration of an ESS shows a positive impact by effectively managing the congestion levels. If we consider the issue of line congestion from the perspective of voltage instability, it becomes evident that in the network without an ESS, the increased load levels during peak hours can lead to voltage violations, as depicted in Fig. \ref{fig:13}. 
However, with an ESS, as shown in Fig. \ref{fig:14}, the voltage profile remains stable and within acceptable limits across all buses. This observation further emphasizes the effectiveness of using ESS to ensure reliable operation of the distribution grids under stressed conditions.

\section{Conclusion} \label{4.conclude}
In this work, we assessed the value of the ESS in distribution grids as the aggregation of several benefits that ESS can provide, namely, cost reduction, energy arbitrage, voltage regulation, and congestion management. We proposed an optimization problem that puts together the DistFlow power flow equations, ESS operation modeling, and optimal ESS sizing and siting. The renewable production from photovoltaic power generators is modeled using representative scenarios. Numerical analyses were performed using a modified IEEE 33-bus system. The results clearly demonstrate the advantages of ESS integration in distribution grids. The ESS effectively mitigates congestion issues, improves grid flexibility, and optimizes power flow. Additionally, it contributes to voltage stability by regulating voltage levels within acceptable limits.
While our research primarily focused on evaluating the value of a single ESS, it is important to acknowledge the potential complementarity and combined benefits of multiple ESS. By concentrating on only one ESS, we aimed to isolate and quantify the cause-effect relationship between an ESS and its impact on the distribution grid. Future research can explore the integration of multiple ESS to realize their synergistic effects, thus providing a more comprehensive understanding of their capabilities in grid optimization. Looking ahead, our future research efforts will be directed towards enhancing solution methodologies to reduce computation time. This will involve the development of innovative approaches, such as decomposition methods and acceleration techniques, to deliver faster and more scalable solutions. By addressing these challenges, we can further advance the integration of the ESS in distribution grids, unlocking their full potential for sustainable and efficient energy management.

\bibliographystyle{elsarticle-num}
\bibliography{refs}

\end{document}